\documentstyle[11pt]{article}

\font\tenrm=cmr10
\font\tenit=cmti10
\font\elevenbf=cmbx10 scaled\magstep 1
\font\elevenrm=cmr10 scaled\magstep 1
\font\elevenit=cmti10 scaled\magstep 1

\renewenvironment{thebibliography}[1]
 { \elevenrm
   \begin{list}{\arabic{enumi}.}
    {\usecounter{enumi} \setlength{\parsep}{0pt}
     \setlength{\itemsep}{3pt} \settowidth{\labelwidth}{#1.}
     \sloppy
    }}{\end{list}}

\def\ee{e^+e^-}
\def\to{\rightarrow}
\def\bbar{{\bar{b}}}
\def\bb{{b\bar{b}}}
\def\tbar{{\bar{t}}}
\def\tt{t\bar{t}}
\def\degree{^{\circ}}

\def\GeV{{\rm GeV}}
\def\cM{{\cal M}}
\def\cR{{\cal R}}
\def\as{\alpha_s}
\def\gt{\Gamma}

\def\kos{1}
\def\dkos{2}
\def\kuhn{3}
\def\jikia{4}

\def\beq{\begin{equation}}
\def\eeq{\end{equation}}

\begin{document}
\vspace*{-1.3cm}
\begin{flushright}
UCD-93-23 \\
hep-ph/9307338 \\
July 1993 \\
\end{flushright}
\vspace*{.5cm}
\begin{center}{{\elevenbf GLUON RADIATION AND TOP WIDTH EFFECTS\footnote{
Work supported in part by the Texas National Research Laboratory
Commission and the United Kingdom Science and Engineering Research Council.
}$^,$\footnote{Presented by L.H. Orr at the Workshop on Physics and Experiments
with Linear $\ee$ Colliders, Waikaloa, HI, April 26-30, 1993.}
\\}
\vglue 0.575cm
{\tenrm LYNNE H. ORR
%\footnote{Address after Sept.\ 1, 1993:  Department
%of Physics and Astronomy, University of Rochester, Rochester, NY 14627.}
\\}
\baselineskip=12pt
{\tenit Department of Physics, University of California\\}
\baselineskip=11pt
{\tenit Davis, CA 95616, USA\\}
\vglue 0.25cm
{\tenrm YU.L. DOKSHITZER \\}
{\tenit Department of Theoretical Physics, University of Lund \\
S\"olvegatan 14A, S-22362 Lund, Sweden \\}
\vglue 0.13cm
{\tenrm and\\}
\vglue 0.13cm
{\tenrm V.A. KHOZE and W.J. STIRLING \\}
{\tenit Department of Physics, University of Durham \\
Durham DH1 3LE, England\\}
\vglue 0.38cm
{\tenrm ABSTRACT}}
\end{center}
{\rightskip=2pc
 \leftskip=2pc
 \tenrm\baselineskip=12pt
\vspace{-.15cm}
 \noindent
The large width of the top quark influences the way gluons are radiated in
top events, giving rise to interesting interference effects in top
production and decay.  We discuss top width effects in
soft guon radiation in $\ee \to \tt$ at high energies and near $\tt$ threshold.
\vglue 0.45cm}
{\elevenbf\noindent 1. Introduction}
\vglue 0.3cm
\baselineskip=13pt
\elevenrm
The mass of the top quark is large enough that top decays to a real $W$ and
a $b$ quark with a width $\gt$ that increases with $m_t$ and
is typically in the GeV range.
This implies a weak lifetime comparable to strong
interaction timescales, which  in turn gives rise to interesting new
effects involving the interplay between top's strong and weak interactions.
For example, a heavy top quark might decay too quickly to form bound
states when it is produced, leaving little or no resonant structure in
the threshold region (which, among other things, makes it difficult to
measure $\gt$).
Here we are interested in the
perturbative side of the strong--weak interplay:  how the top quark's
large decay width influences the way gluons are radiated.
More detailed discussion of the work presented here can be found in
[\kos]\ and [\dkos]; for reviews of theoretical and experimental issues in top
physics, see [\kuhn].

We will discuss soft radiation in $\ee \to \tt$ at high energies and
near $\tt$ threshold, by sketching out
the general formalism and showing some examples of gluon distributions which
are sensitive to $\gt$.  We are motivated in part by the following questions,
which we will return to at the end.
\vspace{-.125cm}
\begin{itemize}
\addtolength{\itemsep}{-4pt}
\item Does the large top width suppress radiation off the top quark, {\it
i.e.,}
does the gluon distribution in top events look as though the
$b$ quarks were produced directly rather than through top decay?
\item Do gluons radiated in the production and decay stages interfere?
\item At/near the $\tt$ threshold, can the top width affect the gluon
distribution, even though the $t$ and $\bar t$ are produced at rest and
do not radiate?
\item What does ``large width'' mean in the context of perturbative
QCD?  In non-perturbative QCD, large means large compared to $\Lambda_{QCD}$.
What is the appropriate dimensionful quantity to which we should
compare $\gt$ here?
\item Can width-dependent effects be exploited to obtain a measurement of $\gt$
from soft gluon distributions?
\end{itemize}
\vglue 0.4cm
{\elevenbf\noindent 2. Soft Gluon Radiation in $\ee \to \tt$: General
Formalism$^{[\kos]}$}
\vglue 0.3cm
Consider $\ee \to \tt \to W^+W^-\bb$ with emission of a (single) gluon.
The gluon can be radiated by any one of the
quarks, and we must include the diagrams for all four possibilities.
In the limit of soft gluons the matrix element $\cM$ factorizes and
we can write $\cM \sim \cM^{(0)} J\cdot\epsilon$,
where $\cM^{(0)}$ is the zeroth-order matrix element and
$J^{\mu}$ and $\epsilon_{\mu}$ are respectively the gluon current and
polarization.
This factorization allows us to define a
gluon emission probability density, which is just the gluon part of the
differential cross section normalized to the zeroth-order cross section
$\sigma_0$, {\it i.e.,}
\beq
dN \equiv 1/\sigma_0 d\sigma_g =
 \frac{d\omega}{\omega}\frac{d\Omega}{4\pi}\>\frac{C_F\as}{\pi}\>\cR\>,
\eeq
where $\omega$ and $\Omega$ denote the gluon energy and solid angle.
$\cR$ is given by integrating the square of the current over the  top-quark
virtualities:
\beq
\cR \equiv  \omega^2 \left({M\Gamma\over\pi}\right)^2 \;
\int dq_1^2 dq_2^2\; [-J\cdot J^*] \; .
\eeq
The full expression for $\cR$ can be found in [\kos]
(see also [\jikia]); we will give some numerical examples below.

The important point is that the current $J^\mu$ can be decomposed
in a gauge-invariant way into terms
that correspond to radiation at the production and decay stages.
Thus, in $\cR$ we can identify unambiguously contributions from gluon
emission in $\tt$ production, $t$ and $\tbar$ decay, and their interference.
$\gt$ enters $J^{\mu}$ via the top quark propagators.  After the integration
in Eq.\ 2, $\gt$ drops out of the pure--production and pure--$t$ (or --$\tbar$)
decay terms and all of
the width dependence is in the production--decay and $t$ decay--$\tbar$ decay
interference.  It is also worth noting that the magnitudes of these
interference
terms increase with increasing $\gt$ and vanish as $\gt\to0$.

It will be useful to keep in mind two limiting cases.
Our intuition tells us that if the top width is large, the top
quark decays before it has time to radiate, and the gluon distribution
is identical to that for a $b$ and $\bbar$ produced directly
(assuming the same final state kinematics);
as $\gt\to\infty$, the interference cancels all contributions that
involve emission off the top quark lines.  On the other hand, if the width is
small, the interference is small and the radiation in the
production and decay stages can be treated independently.\footnote{Thus
the way the top quark is treated in Monte Carlo simulations usually
corresponds to one or the other limiting value of $\gt$.}

\begin{figure}
\vspace{14cm}
\vspace{7cm}
\hspace{-3.2cm}
\hspace{-.2in}
\special{insert [orr.corsica.papers.figures]lcws1.ps}
\vspace{-14.95cm}
\caption{Soft gluon distribution in $e^+e^-\to t \tbar$ for
c.m.\ energy 1 TeV, $m_t=140\ \GeV$,
 $\omega = 5\ \GeV$ and $\phi=0\degree$.  $\theta$ is the $t$--$g$ angle.
(a) $\theta_b=180\degree.$ (b) $\theta_b=90\degree.$}
\vspace{-.23cm}
\end{figure}

\vglue 0.4cm
{\elevenbf \noindent 3. Collisions at High Energies$^{[\kos]}$}
\vglue 0.3cm

At high collision energies, gluons radiated at the top
production and decay stages interfere destructively for most configurations,
and the decay--decay interference is negligible.
Thus the effect of the top width is to suppress the gluon distribution.
In addition, it turns out that the inteference is largest for large $b$--$t$
angular separations.

This is illustrated in Figure 1, where we show the gluon emission
probability as
a function the angle $\theta$ of the gluon with respect to the top-quark
direction.  We vary the top width, and
take $m_t=140$ GeV, collision energy 1 TeV, and gluon
energy $\omega=5$ GeV.  Because in the soft limit we neglect the gluon momentum
in the kinematics, the $t$ and $\tbar$ are produced back--to--back.  For
simplicity we also choose the $b$ and $\bbar$ back--to--back.
In Fig.\ 1(a)
$\theta_b=180\degree$, that is, the $t$ decays to a
backward $b$.  The different curves correspond to different values of the top
width $\gt$.
As $\gt$ increases, the peaks in the gluon distribution are
suppressed and eventually disappear.  We see also that for the SM width
($\gt=0.7$\ GeV at lowest order\footnote{$\gt$ appears as a parameter here, and
so using the lowest order value is adequate for our purposes.  Corrections to
the top width are discussed elsewhere.$^{[\kuhn]}$})
the gluon distribution is much closer to the
zero-width curve than that for $\gt=\infty$.

Similar behavior appears in Fig.\ 1(b), where we show the slightly
more likely configuration where the $b$'s come off at right
angles from their parent $t$'s ($\theta_b=90\degree$).
There is more structure in the distributions because this configuration
is less symmetric than in Fig.\ 1(a).  As a result, the gluon distribution
not only gets suppressed but also changes shape as the width increases
and wipes out all traces of the top quarks at $\theta=0\degree$ and
$180\degree$.

Not shown is the case where $\theta_b=0\degree$; while it is by far the most
likely configuration (because a high energy top quark tends to decay to a
collinear $b$), there is very little interference and the width has
virtually no effect on the distribution.  That this must be true becomes
clear when we think in terms of moving color charges: QCD is flavor-blind, so
the decay entails only a slight change in speed, but not direction, of the
color source.

\begin{figure}
\vspace{14cm}
\vspace{7cm}
\hspace{-3.2cm}
\special{insert [orr.corsica.papers.figures]lcws2.ps}
\vspace{-15.65cm}
\caption{Soft gluon distribution in $e^+e^-\to t \tbar$ for
$m_t=140\ \GeV$,
near $\tt$ threshold, with gluon perpendicular to $\bb$ plane;
$\theta_{12}$ is the $b$--$\bbar$ angle.}
\vspace{-.28cm}
\end{figure}

\vglue 0.4cm
{\elevenbf \noindent 4. Collisions Near $\tt$ Threshold$^{[\dkos]}$ \hfil}
\vglue 0.3cm

It is perhaps more interesting to consider the $\tt$ threshold region,
which is of particular importance to top physics.$^{[\kuhn]}$
One may ask whether the size of the width influences the
radiation of gluons for top pairs produced near threshold.
It might seem unlikely, because the $t$ and $\tbar$ are produced nearly at rest
and only the $b$'s radiate.  However, the top width {\it does} affect
the distribution by determining to what extent the radiation from the $b$'s is
coherent or independent.  This can be understood by considering the same
limiting cases as above.  As $\gt\to\infty$, the top lifetime
approaches $0$, the $b$ and $\bbar$ appear instantaneously, and they radiate
coherently.  But if $\gt$ is very small, the top lifetime is large, the $b$
and $\bbar$ appear at very different times and they radiate independently,
with no interference.
Thus $\gt$ regulates the interference
between gluons radiated by the $b$ and $\bbar$ and, again, in the limit of
large top width the radiation looks as though the $b$ and $\bbar$ were produced
directly.

This can be made quantitative by taking the threshold limit of the expression
for $\cR$ in Eq.\ 2.\footnote{The same result can be obtained
from a semi-classical derivation based on the motion of color sources
and the quantum-mechanical arguments of the previous paragraph; see
[\dkos].}
If $v$ is the velocity of the $b$
(or $\bbar$), $\theta_{1(2)}$ is the angle between the $b$ ($\bbar$)
and the gluon, and $\theta_{12}$ is the angle between the $b$ and $\bbar$, then
we obtain
\beq
\cR = \frac{v^2\sin^2\theta_1}{(1-v\cos\theta_1)^2}\> +\>
\frac{v^2\sin^2\theta_2}{(1-v\cos\theta_2)^2}\>
+\>
2 \chi\,
\frac{v^2(\cos\theta_1\cos\theta_2-\cos\theta_{12})}
{(1-v\cos\theta_1)(1-v\cos\theta_2)},
\eeq
where $\chi \equiv { \Gamma^2 \over \Gamma^2 + \omega^2}$.
The first two terms correspond to independent emission by the $b$ and $\bbar$,
and the last term is the interference, regulated by the
top width through $\chi$.  Note that $0\leq\chi\leq1$
and the extreme values of $\chi$ correspond to the limits of independent
($\gt=0$) and coherent ($\gt\to\infty$) emission.
Note also that $\chi$ depends not on $\Gamma$ alone but on its ratio to the
gluon energy $\omega$; hence $\omega$ determines the relevant scale for
$\gt$.
Furthermore, from the form of $\chi$ it is clear that we have maximum
sensitivity to the top width when $\gt$ and $\omega$ are comparable.

\begin{figure}
\vspace{14cm}
\vspace{7cm}
\hspace{-3.2cm}
\hspace{-.2in}
\special{insert [orr.corsica.papers.figures]lcws3.ps}
\vspace{-15.5cm}
\vspace{.25in}
\caption{Soft gluon emission probablilty near $\tt$ threshold
($m_t=140\ \GeV$)
integrated over $5 \le\omega\le 10$\ GeV, $0\le\phi\le360\degree$,
and (a) $0\le\theta\le\theta_{12}$ and
(b) $\theta_{12}\le 180\degree$;
angles are measured from the $b$ direction.
$\Gamma=0.7,\ 3,\ 5\ \GeV$ for the dotted, dashed, and dashed--dotted lines.}
\vspace{-.23cm}
\end{figure}

Let us look at some examples.
Since the interference term contains the width dependence, and it is also the
only term that depends on the relative orientation
of the $b$ and $\bbar$, it will be useful to consider
how the gluon distributions vary with $\theta_{12}$.
A particularly simple radiation pattern is that for gluons emitted
perpendicular to the $\bb$ plane, in which case  $\theta_1=\theta_2=\pi/2$
and $\cR \propto (1-\chi\cos\theta_{12})$.  Figure 2 shows the distribution
for various values of $\chi$.  The $\chi=0$ distribution is flat
and exhibits no interference effects.  As $\chi$ increases,
$\theta_{12}$ dependence is induced and we see that the interference
is destructive for small $\theta_{12}$ and constructive for large
$\theta_{12}$.

For a 5 GeV gluon, our 140 GeV top quark with a
width of 0.7 GeV gives $\chi\approx 0.02$,
much closer to the independent emission case than the large width limit.
If the width were about as large as $\omega$, or,
conversely, if we could detect gluons with energies of about 1 GeV,
we would see that the radiation pattern matched neither the large nor
the small width limit, and that the distribution was very sensitive to the
exact value of $\gt$.

Interference between gluons gives rise to so-called angular ordering behavior:
After integration over the azimuthal angle about the $b$ quark direction,
emission is enhanced between the $b$ and $\bbar$ and suppressed outside them.
Here \lq\lq between'' and ``outside'' refer to the gluon polar angle $\theta$
(with respect to the $b$ direction) respectively less or greater
than $\theta_{12}$.  This effect can be seen in Figure 3, where we show
(partly integrated) distributions for gluons (a) between and (b) outside
the $\bb$ pair for $\gt$ ranging from $0$ to $\infty$.

Finally, we show in Figure 4 the total gluon distribution as a function of
$\theta_{12}$, integrated over all angles and
energies from $5$ to $10$ GeV.  We see, again, no $\theta_{12}$ dependence
for $\gt=0$, and, as $\gt$ increases, the interference --- destructive
for small $\bb$ angles and constructive for large $\bb$ angles --- changes
the shape of the curve.  As in the differential distributions, the SM case
($\gt=0.7$) GeV gives a result much closer to that for $\gt=0$ than infinity.
Because the top width is so {\it small}, interference
between the $b$ and $\bbar$ is almost completely suppressed, contrary to
the large-width expectation of coherent emission as if the $b$ and $\bbar$
were produced directly.

\begin{figure}
\vspace{14cm}
\vspace{7cm}
\hspace{-3.2cm}
\special{insert [orr.corsica.papers.figures]lcws4.ps}
\vspace{-15.5cm}
\caption{Soft gluon emission probablilty near $\tt$ threshold
($m_t=140\ \GeV$)
integrated over all gluon angles and energies
from 5 to 10 GeV.}
\vspace{-.23cm}
\end{figure}

\vglue 0.4cm
{\elevenbf \noindent 5. Conclusions \hfil}
\vglue 0.3cm

We conclude by answering explicitly the questions raised in the
Introduction.
\begin{itemize}
\addtolength{\itemsep}{-4pt}
\item {\it Does the top width suppress radiation off the top quark?}
No;  unless $\gt$ is extremely large, $t$ and $\tbar$ emission
contribute substantially to the gluon distribution.
\item {\it Do gluons radiated in the production and decay stages interfere?}
Yes, and the interference is destructive at high energies, leading
to a suppression of radiation and sometimes a change in the shape of
the distribution.
\item {\it At/near the $\tt$ threshold, can the top width affect the gluon
distribution?}  Yes; the top width suppresses the interference between the
$b$ and $\bbar$ by a factor $\chi \equiv { \Gamma^2 \over \Gamma^2 +
\omega^2}$.
\item {\it What does ``large width'' mean in the context of perturbative
QCD, i.e., what is the appropriate quantity to which we should
compare $\gt$?}  The gluon energy $\omega$ sets the scale for
the width.  As we have seen, although
a width of 0.7 GeV is large compared to $\Lambda_{QCD}$, it is not large
compared to accessible gluon energies.  Therefore, contrary to what we might
expect, a 140 GeV top quark has a small width for PQCD purposes.
\item {\it Can width-dependent effects be exploited to measure $\gt$
from soft gluon distributions?}
There is maximum
sensitivity to the width when it is comparable to the gluon
energy.  If the minimum accessible gluon energy is around $5$ GeV,
then the best prospects for a measurement are for $\gt$ also a few
GeV or higher, {\it i.e.,} for an anomalously large width (for \lq\lq typical''
top masses around 140-150 GeV), or for a very heavy top quark.
In general, the larger the top width, the better the
prospects for using soft gluon radiation to measure it.
\end{itemize}

\vglue 0.4cm
{\elevenbf\noindent 6. References \hfil}
\vglue 0.3cm

\end{document}